\documentclass[prl,preprint,showpacs]{revtex4}
\usepackage{amsmath,amsfonts}
\usepackage{epsfig}

\begin{document} 
%%%%%%%%%%%%%%%%%%%%%%%%%%%%%%%%%%%%%%%%%%%%%%%%%%%%%%%%%%%%%%%%%%%%%%%%%%%%
%%%%%%%%%%%%%%%%%%%%%%%%%%%%%%%%%%%%%%%%%%%%%%%%%%%%%%%%%%%%%%%%%%%%%%%%%%%%

\begin{flushright}
CP3-08-18\\
ICMPA-MPA/2008/13\\
NITheP-08-01\\
October 2008
\end{flushright}

\title{Thermodynamics of a non-commutative fermion gas}  
\date{\today}
\author{Frederik G. Scholtz$^{a,b}$\footnote{E-mail: fgs@sun.ac.za} and Jan Govaerts$^{b,c,d}$\footnote{Fellow of
the Stellenbosch Institute for Advanced Study (STIAS), Stellenbosch, Republic of South Africa.\\E-mail: Jan.Govaerts@uclouvain.be}}
\affiliation{$^a$Institute of Theoretical Physics, University of Stellenbosch, Stellenbosch 7600, South Africa\\
$^b$National Institute for Theoretical Physics (NITheP), Stellenbosch Institute for Advanced Study (STIAS), 
7600 Stellenbosch, South Africa\\
$^c$Center for Particle Physics and Phenomenology (CP3),
Institut de Physique Nucl\'eaire, Universit\'e catholique de Louvain (U.C.L.),
2, Chemin du Cyclotron, B-1348 Louvain-la-Neuve, Belgium\\
$^d$International UNESCO Chair in Mathematical Physics and Applications (ICMPA),
University of Abomey-Calavi, 072 B.P. 50, Cotonou, Republic of Benin}

\begin{abstract}
Building on the recent solution for the spectrum of the non-commutative well \cite{fgs} in two dimensions, the thermodynamics that follows from it is computed. In particular the focus is put on an ideal fermion gas confined to such a well. At low densities the thermodynamics is the same as for the commutative gas.  However, at high densities the thermodynamics deviate strongly from the commutative gas due to the implied excluded area resulting from the non-commutativity.  In particular there are extremal macroscopic states, characterized by area, number of particles and angular momentum, that correspond to a single microscopic state and thus have vanishing entropy.  When the system size and excluded area are comparable, thermodynamic quantities, such as entropy, exhibit non-extensive features. 
   
\end{abstract}
\pacs{11.10.Nx}

\maketitle

%%%%%%%%%%%%%%%%%%%%%%%%%%%%%%%%%%%%%%%%%%%%%%%%%%%%%%%%%%%%%%%%%%%%%%%%%%%%
There seems to be growing consensus that our notion of space-time has to be drastically revised in a consistent formulation of quantum mechanics and gravity \cite{dop,seib}.  One possible generalization, suggested by string theory \cite{wit}, that has attracted much interest recently is that of non-commutative space-time \cite{doug}.  Despite a number of investigations into the possible physical consequences of non-commutativity in quantum mechanics and quantum mechanical many-body systems \cite{li,mendes,bem,khan}, quantum electrodynamics \cite{chai,chair,lia}, the standard model \cite{ohl} and cosmology  \cite{gar,alex}), our understanding of the physical implications of non-commutativity is still in its infancy. In particular our understanding of the physics is cluttered by the implementation of the Poincar\'e symmetry.  One either has to accept explicit breaking of the Poincar\'e symmetry, with its possible physical ramifications in e.g. high energy cosmic radiation \cite{bret}, or one can restore the Poincar\'e invariance by twisting the Poincar\'e group \cite{chai1,wess1,asch,bal,chak1}.  In the latter case it is also necessary to twist the statistics, but there is a controversy on whether this should be done in the sense of \cite{bal,chak1}, which leads to modified statistical correlations \cite{chak1}, or the braided sense of \cite{wess,fiore} and it seems difficult to distinguish between these two possibilities.  Indeed, the physical effects of non-commutative space-time and modified statistical correlations arising from twisted statistics seem to be  irrevocably intertwined. Even in the case of a non-interacting, but confined non-commutative gas, the physical consequences of non-commutativity have not been investigated in detail due to the problem of giving precise meaning to boundaries in a non-commutative space. Against this backdrop it is highly desirable to find systems where the physical effects of non-commutativity can be investigated exactly and unambigously.

Recently precise meaning was given to a particle confined to a two dimensional non-commutative infinite well \cite{fgs}, the spectrum was computed exactly and it was shown that the results reduce to the conventional results in the commutative limit. The result, summarized below, is that at low energies there is very little distinction between the commutative and non-commutative world, if the non-commutative parameter is small compared to any other length scales squared in the system, which will be the case if the non-commutative parameter is of the order of the Planck length squared. Thus one expects that at normal densities (interparticle spacings) and temperatures (thermal length) non-commutativity should be virtually undetectable.  At high densities or temperatures this situation may, however, change.  This motivated us to study the thermodynamics of such systems at extreme densities and pressures, which is what we report on here.  The availability of exact results for the spectrum enabled us to study these limits reliably, albeit numerically.  Furthermore in two dimensions the commutator $\left[x,y\right]$ is already a $SO(2)$ invariant ($\left[x^\prime,y^\prime\right]=\left[x,y\right]$) as one can easily check and thus the non-commutative relation $\left[x,y\right]=i\theta$, where $\theta$ is considered to be a $SO(2)$ invariant, does not break the rotational symmetry.  In two dimensions it is therefore not necessary to restore the rotational symmetry by twisting the rotation group.  This is in contrast to the higher dimensional case where the non-commutative relation $\left[x_i,x_j\right]=i\theta_{ij}$ breaks the rotational (Poincar\'e) invariance if the $\theta_{ij}$ are considered to be fundamental invariants which are the same for all observers and a twist of the rotational symmetry is required to restore the rotational invariance. The complication of a twisted rotational symmetry, twisted statistics and possible modified statistical correlations arising from it does therefore not occur in two dimensions, which opens up the possibility of studying the physical effects of non-commutativity in an unambiguous setting.  There is a possible exception to this in the case of a semi-direct product of a twisted gauge symmetry and rotational symmetry \cite{bal1}, but this does not apply to our current investigation.  Furthermore it should be pointed out that, apart from the issue of the restoration of rotational symmetry and the associated question of statistics, the current analysis can be extended straightforwardly to higher dimensions. Indeed, in three dimensions one can, by an appropriate change of coordinates, always reduce the non-commutativity to two non-commutative and one commutative coordinate. As expected this breaks rotational invariance, but if one does not insist on its restoration, the analysis given below extends with minor modifications to the higher dimensional case.  This, and the associated issue of the restoration of rotational symmetry, will, however, be the topic of future investigations.

We start by summarizing the key results of \cite{fgs} which we use here.  The spectrum of a particle with mass $m_0$ confined to a disc of radius $R^2=\theta(2M+1)$, with $[x,y]=i\theta$ ($\theta>0$), in a non-commutative space is given by    
\begin{eqnarray}
\label{infwellnc}
L_{M+1}^m\left(\frac{\theta k^2}{2}\right)&=&0,\quad m\ge 0\nonumber\\
L_{M-|m|+1}^{|m|}\left(\frac{\theta k^2}{2}\right)&=&0,\quad -M\le m< 0,\\
E&=&\frac{\hbar^2k^2}{2m_0}\nonumber.
\end{eqnarray}
Here $M$ is an arbitrary non-negative integer that fixes the radius of the disc, $m$ is the angular momentum and $L_n^m$ an associated Laguerre polynomial. Eigenfunctions for $m<-M$ vanish identically and the spectrum truncates at angular momentum $m=-M$.  A crucial observation, underlying many of the features discussed below, is that each positive angular momentum sector has exactly $M+1$ excitations and each negative angular momentum sector, $-M\le m<0$, exactly $M-|m|+1$ excitations.  For later reference we give a graphical summary of these results in Fig.~1.  
\setlength{\unitlength}{1mm}
\begin{figure}
\begin{picture}(53,53)
\put(-20, 0){\epsfig{file=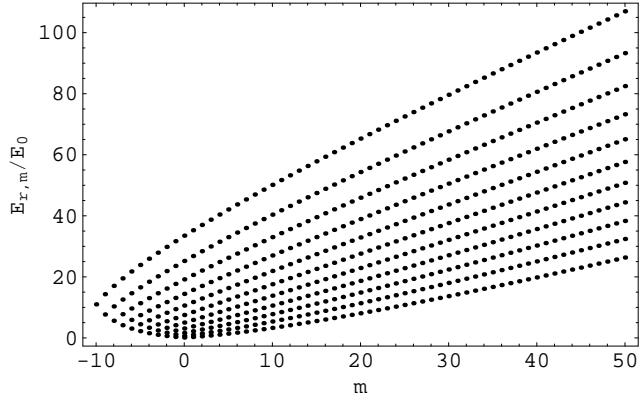, height=53mm}}
\end{picture}
\caption {Spectrum of a particle in a non-commutative well for $M=10$ in units of $E_0=\frac{\hbar^2}{m_0\theta}$ as a function of angular momentum.}
\end{figure}

The spectrum for the corresponding commutative system is given by
\begin{eqnarray}
\label{infwellc}
J_m\left(kR\right)&=&0,\quad m=0,\pm 1,\pm 2\ldots\nonumber\\
E&=&\frac{\hbar^2k^2}{2m_0}.
\end{eqnarray}
with $J_m$ a Bessel function.

To compute the spectrum and thermodynamics of the system, we must therefore be able to compute the zeros of the Laguerre polynomials and Bessel functions.  While efficient algorithms to determine the zeros of the Bessel functions are available, the most efficient and accurate determination of the zeros of the Laguerre polynomials is to find the eigenvalues of the following $N\times N$ hermitian matrix, constructed such that its eigenvalues yield exactly the zeros $x_{r,m}$ ($1\le r\le N$) of the Laguerre polynomial $L_N^m(x)$:
\begin{equation}
H_{i,j}(N,m)=\left(2i-1+m\right)\delta_{i,j}+\sqrt{i(i+m)}\delta_{j,i+1}+\sqrt{j(j+m)}\delta_{j,i-1},\quad i,j=1,2,\ldots N.
\end{equation}
Indeed, the characteristic polynomial in $x$ of this matrix obeys a three term recurrence relation which is that of the
Laguerre polynomial $L^m_N(x)$ up to a normalisation factor, ${\rm det}\left(H(N,m)-x\mathbb{I}\right)=N!\,L^m_N(x)$.

From Fig.~1 it should be clear that for certain choices of total number of particles and total angular momentum, the single particle states must be filled in a unique way (for modelling purposes in the present paper, these particles are taken to obey a Fermi statistics with a single spin state).  Indeed, let us consider the situation where all the single particle states from angular momentum $-M$ up to angular momentum $m$ are completely filled. The total number of particles is
\begin{eqnarray}
\label{parnum}
N_m&=&\sum_{k=0}^{m+M} (k+1)=\left(1 + m + M\right)\left(1 + \frac{1}{2}\left( m + M \right)\right),\;-M\leq m<0,\nonumber\\
&=&\sum_{k=0}^{M-1}(k+1)+(m+1)(M+1)=\frac{1}{2}\left( 1 + M \right) \,\left( 2 + 2\,m + M \right),\;m\geq 0.
\end{eqnarray}     
The total angular momentum of the state is
\begin{eqnarray}
\label{minangmom}
L_{m}&=&\sum_{k=0}^{m+M}(k+1)(k-M)=\frac{1}{6}\left( 2\,m - M \right) \,\left( 1 + m + M \right) \,\left( 2 + m + M \right),\;-M\leq m<0,\nonumber\\
&=&\sum_{k=0}^{M-1} (k+1)(k-M)+\sum_{k=0}^m(M+1)k \nonumber\\
&=&\frac{1}{6}\left( 1 + M \right) \,\Big( 3\,m\,\left( 1 + m \right)  - M\,\left( 2 + M \right)  \Big),\;m\geq 0.
\end{eqnarray}
It is clear that any other distribution of the number of particles (\ref{parnum}) must lead to a higher angular momentum than (\ref{minangmom}), which is thus the minimum total angular momentum for this number of particles and system size $M$.  This implies that if we fix, for system size $M$, the total angular momentum on (\ref{minangmom}) and the number of particles on (\ref{parnum}) for some $m$, there is only one possible distribution of particles.  The entropy for this macroscopic state thus vanishes.  Furthermore it is clear that increasing the particle number, while keeping the system size fixed, or decreasing the system size while keeping the particle number fixed, must yield angular momentum values larger than $L_{m}$. Thus, if the total angular momentum, which is a conserved quantum number, is fixed on $L_{m}$, we expect these states to be incompressible and the pressure to diverge as either the particle number or system size approach the critical values from below or above, respectively. Alternatively we expect the pressure to diverge as the particle density approaches the critical value $\rho_m=\frac{N_m}{\pi\theta(2M+1)}$ from below.  The total energy of these states can also be evaluated quite easily and in units of $E_0=\frac{\hbar^2}{m_0\theta}$ is given by,
\begin{eqnarray}
\frac{E_m}{E_0}&=&\sum_{k=0}^{m+M}{\rm tr}H(k+1,M-k)\nonumber\\
&=&\left( 1 + M \right) \,\left( 1 + m + M \right) \,\left(1 + \frac{1}{2}\left( m + M \right) \right),\;-M\leq m<0,\nonumber\\
&=&\sum_{k=0}^{M-1}{\rm tr}H(k+1,M-k)+\sum_{k=0}^{m}{\rm tr}H(M+1,k)\nonumber\\
&=&\frac{1}{2}\left( 1 + M \right) \,\Big( M\,\left( 1 + M \right)  + \left( 1 + m \right) \,\left( 2 + m + 2\,M \right)  \Big),\;m\geq 0.
\end{eqnarray}        
 
For the extremal cases above we have considered the micro canonical ensemble, i.e., fixed area, particle number, total angular momentum and energy. It is, however, only in these extreme cases that that the computation of the micro canonical partition function and thermodynamics is trivial.  For general particle numbers and angular momenta this becomes very difficult and to proceed in these cases it is more appropriate to switch to the grand canonical ensemble in which the average particle number, average energy and average total angular momentum are fixed.  The reason for fixing the average total angular momentum is to ensure that when the commutative and non-commutative systems are compared, it is done for the same physical situation.  It should be clear from the spectra of the commutative and non-commutative systems that fixing only the particle number and temperature may lead to vastly different average total angular momenta.

The appropriate density matrix for this ensemble that acts on Fermi Fock space is
\begin{eqnarray}
\rho&=&{\cal L}^{-1}e^{-\beta(\hat H-\mu \hat N-\omega \hat L)},\nonumber\\
{\cal L}&=&{\rm tr}e^{-\beta(\hat H-\mu \hat N-\omega \hat L)}.
\end{eqnarray}
Here the Hamiltonian, $\hat H$, particle number operator, $\hat N$, and total angular momentum, $\hat L$, are given in second quantized form by
\begin{eqnarray}
\hat H&=&\sum_{r,m}E_{r,m}a^\dagger_{r,m}a_{r,m},\nonumber\\
\hat N&=&\sum_{r,m}a^\dagger_{r,m}a_{r,m},\\
\hat L&=&\sum_{r,m}\hbar m a^\dagger_{r,m}a_{r,m},\nonumber
\end{eqnarray}
where $a^\dagger_{r,m}$, $a_{r,m}$ creates (annihilates) Fermi particles with energy $E_{r,m}$ and angular momentum $m$. 
Furthermore $\beta=1/kT$, $\mu$ is the chemical potential and $\omega$ a Lagrange multiplier that enforces the constraint of fixed total angular momentum, which, for rigid bodies, has the physical meaning of the angular velocity.  The partition function, ${\cal L}$, which involves a trace over full Fermi Fock space, is explicitly given by 
\begin{equation}
{\cal L}={\rm tr}e^{-\beta(\hat H-\mu \hat N-\omega \hat L)}=\prod_{r,m}\left(1+e^{-\beta(E_{r,m}-\mu-\hbar\omega m)}\right).
\end{equation}
The central thermodynamic quantity from which all thermodynamics derive, is the q-potential which is the logarithm of the partition function
\begin{equation}
q(A,T,\mu,\omega)=\log{\cal L}=\sum_{r,m}\log\left(1+e^{-\beta(E_{r,m}-\mu-\hbar\omega m)}\right).
\end{equation}
The q-potential depends on the temperature, chemical potential, angular velocity and area, $A$, of the system.  The dependence on area enters through the single particle energies that depend on it.
 
To find the thermodynamic interpretation of the q-potential one can proceed as in \cite{pathria} p.101 with slight modification due to the constraint on angular momentum.  For completeness we briefly outline the argument.  Taking the differential of the q-potential one has
\begin{equation}
dq=-d\beta\left(E-\mu N-\omega L\right)-\beta\langle\frac{d\hat H}{dA}\rangle dA+\beta Nd\mu+\beta Ld\omega
\end{equation}
where $E$, $N$, $L$ and $\langle\frac{d\hat H}{dA}\rangle$ denote the average energy, particle number, angular momentum and average of $\frac{d\hat H}{dA}$, respectively.  This can be rewritten as 
\begin{equation}
d\left(q+\beta\left(E-\mu N-\omega L\right)\right)=\beta\left(dE-\mu dN-\omega dL-\langle\frac{d\hat H}{dA}\rangle dA\right).
\end{equation}
Comparing the right hand side with the first law for a rotating body \cite{lifs} p.74
\begin{equation}
TdS=dE+PdA-\mu dN-\omega dL,
\end{equation}
one concludes that $P=-\langle\frac{d\hat H}{dA}\rangle$ and 
\begin{equation}
d\left(q+\beta\left(E-\mu N-\omega L\right)\right)=\frac{dS}{k},
\end{equation}
or,
\begin{equation}
q=\frac{S}{k}-\beta\left(E-\mu N-\omega L\right).
\end{equation}
Keeping in mind that the Gibbs free energy for a rotating body is given by \cite{lifs} p.74
\begin{equation}
G=E-\omega L-TS+PA=\mu N,
\end{equation}
we have
\begin{equation}
q=\frac{PA}{kT}.
\end{equation}
The rest of the thermodynamics follow easily with the quantities most important to our analysis given by
\begin{eqnarray}
\label{threl}
N&=&kT\frac{\partial q}{\partial\mu},\nonumber\\
L&=&kT\frac{\partial q}{\partial\omega},\\
S&=&k\frac{\partial(Tq)}{\partial T}.\nonumber
\end{eqnarray}

We are now in a position to systematically compute the thermodynamics of the commutative and non-commutative Fermi gas. To do this, and for comparison, it is convenient to first rewrite the expression for the q-potential in terms of dimensionless quantities. We start with the non-commutative case.  Let $x_{r,m}$ denote the zeros of the Laguerre polynomials $L_{M+1}^m\left(x_{r,m}\right)=0$. From (\ref{infwellnc}) we then have $E_{r,m}=\frac{\hbar^2 x_{r,m}}{\theta m_0}$. Introducing the energy scale $E_0=\frac{\hbar^2}{\theta m_0}$ and the dimensionless parameters $\tilde\beta=E_0\beta$, $\tilde\mu=\frac{\mu}{E_0}$ and $\tilde\omega=\frac{\hbar\omega}{E_0}$, the q-potential for the non-commutative well can explicitly be written as
\begin{eqnarray}
q(M,\tilde\beta,\tilde\mu,\tilde\omega)=\sum_{m=-M}^\infty\sum_r\log\left(1+e^{-\tilde\beta(x_{r,m}-\tilde\mu-\tilde\omega m)}\right).
\end{eqnarray}
The dependence on the area of the system is reflected by the dependence on $M$, keeping in mind the relation $A=\pi R^2=\pi\theta(2M+1)$.

The commutative case is treated in a similar fashion. We denote by $j_{r,m}$ the zeros of the Bessel functions $J_m\left(j_{r,m}\right)=0$.  From (\ref{infwellc}) we then have $E_{r,m}=\frac{\hbar^2 j_{r,m}^2}{2m_0R^2}$. 
 Using the relation $R^2=\theta(2M+1)$, the q-potential for the commutative well can be expressed as
\begin{eqnarray}
q(M,\tilde\beta,\tilde\mu,\tilde\omega)=\sum_{r,m}\log\left(1+e^{-\tilde\beta(\frac{j_{r,m}^2}{4M+2}-\tilde\mu-\tilde\omega m)}\right).
\end{eqnarray}

In terms of these dimensionless variables the thermodynamic relations (\ref{threl}) read
\begin{eqnarray}
\label{threld}
N&=&\frac{1}{\tilde\beta}\frac{\partial q}{\partial\tilde\mu},\nonumber\\
\frac{L}{\hbar}&=&\frac{1}{\tilde\beta}\frac{\partial q}{\partial\tilde\omega},\nonumber\\
\frac{S}{k}&=&q-\tilde\beta\frac{\partial q}{\partial\tilde\beta}.
\end{eqnarray}

We start our analysis by noting a fundamental difference between the commutative and non-commutative systems.  In the discussion of the extremal states, we have observed that there is a minimum total angular momentum for a given system size and particle number.  Thus, for a fixed angular momentum and system size, $M$, the number of particles in the non-commutative system can not be made arbitrarily large.  When all the negative angular momentum states are filled, the angular momentum must necessarily increase with particle number until the minimum total angular momentum will exceed the given total angular momentum value.  This is shown in Fig. 2 where the number of particles is shown as a function of the chemical potential for fixed angular momentum $L=0$.  One observes that the particle number saturates at a maximum value.  This implies that the non-commutative system has, for a given angular momentum and system size, a maximal density, while no such limit exists in the commutative system. This cut-off in the density for the non-commutative system is a reflection of the implied excluded area resulting from non-commutativity.  The maximal density clearly has a temperature dependence, which turns out to be very weak at the temperatures we consider and we shall neglect it in what follows.  Note that, not surprisingly, the results for the commutative and non-commutative cases agree at low densities, but that they start diverging strongly at high densities.   
\setlength{\unitlength}{1mm}
\begin{figure}
\begin{picture}(53,53)
\put(-20, 0){\epsfig{file=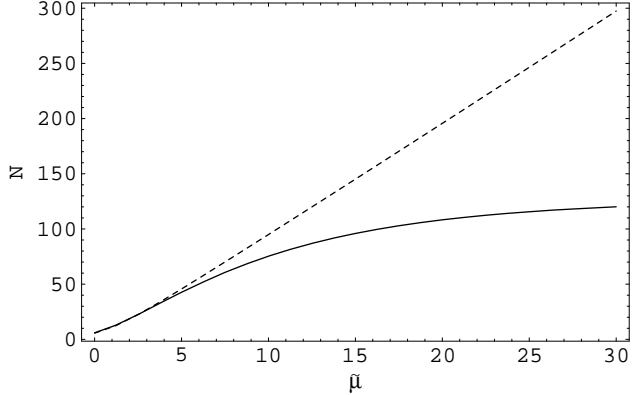, height=53mm}}
\end{picture}
\caption {Average number of particles as a function of the chemical potential at temperature $\tilde\beta=1$, system size $M=10$ and average angular momentum $L=0$.  The solid line is the non-commutative case and the dashed line the commutative case.}
\end{figure}

The maximal particle density as a function of system size can easily be computed and is depicted in Fig. 3. Here we show the dimensionless particle density $\tilde\rho$ defined by $\rho=\frac{N}{\pi R^2}=\frac{N}{\pi\theta(2M+1)}\equiv \frac{\tilde\rho}{\pi\theta}$. The shaded region is forbidden, i.e., each system size has a maximal density associated with it.  Conversely, a fixed density can only be accommodated by systems with a size greater than a minimal size.
\setlength{\unitlength}{1mm}
\begin{figure}
\begin{picture}(65,65)
\put(0, 0){\epsfig{file=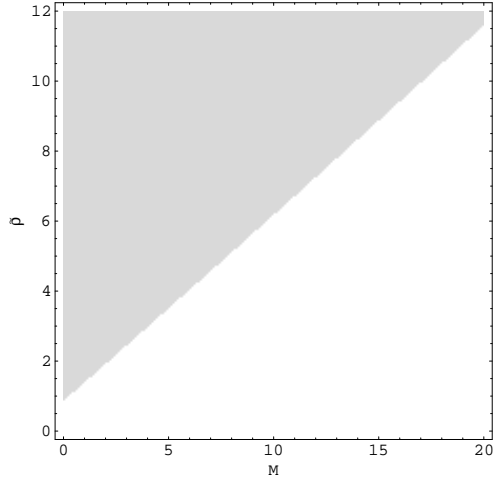, height=65mm}}
\end{picture}
\caption {Maximal dimensionless density as function of system size at $\tilde\beta=1$ and $L=0$. The unshaded region is physical.}
\end{figure}
We can also find an analytic estimate for the curve shown in Fig.~3.  From (\ref{minangmom}) we can easily establish the value of the angular momentum, $m$, up to which all levels must be filled in order to yield $L=0$.  In general this is not an extremal situation, but it is simple to see that this only introduces correction of higher order in $1/M$. Thus, to leading order in $M$ we find $m\approx \frac{M}{\sqrt{3}}$.  Substituting this in (\ref{parnum}) we find for the dimensionless density $\tilde\rho\approx \frac{1}{4}\left(1+\frac{2}{\sqrt{3}}\right)M=0.539 M$. The corresponding slope in Fig.~3 is 0.536. Although our analytic estimate of this slope is done at zero temperature, while the computation in Fig.~3 is done at $\tilde\beta=1$, we still expect good agreement as $\tilde\beta=1$ is of the same order or less than the dimensionless single particle level spacing, i.e., we are approximately at zero temperature.  This, combined with the weak temperature dependence of the maximal density, lead us to expect that this temperature difference will have a small effect, as is indeed the case.  

Next we consider what happens in the non-commutative system as the maximal density is approached.  As remarked before, we expect on purely physical grounds an incompressible behaviour with diverging pressure.  In Fig. 4 we compute the pressure as a function of particle density for a fixed system size, i.e, the pressure as a function of the number of particles. We observe that for the non-commutative system the pressure does indeed diverge at the maximal density, also shown on the graph, while the pressure for the commutative system shows no anomalous behaviour.  The pressure shown is a dimensionless pressure defined by 
\begin{eqnarray}
\tilde P=\frac{\pi\theta P}{E_0}=\frac{q}{\tilde\beta(2M+1)}.
\end{eqnarray}
We note that at low densities the commutative and non-commutative equations of state coincide, while they deviate strongly at high densities.     
\setlength{\unitlength}{1mm}
\begin{figure}
\begin{picture}(53,53)
\put(-20, 0){\epsfig{file=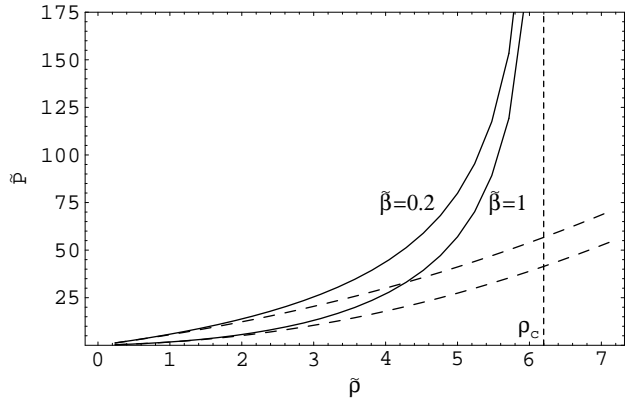, height=53mm}}
\end{picture}
\caption {Dimensionless pressure as a function of density for the non-commutative (solid line) and commutative (dashed line) systems at two temperatures $\tilde\beta=0.2$ and $\tilde\beta=1$, system size $M=10$ and angular momentum $L=0$.  The critical density at $\tilde\beta=1$, where the non-commutative pressure diverges, is indicated by the vertical dashed line.  As remarked in the text, the critical density has a weak temperature dependence and its values virtually coincide for the two temperatures shown, thus we neglect this difference here.}
\end{figure}

Let us consider the dependence of the pressure on the system size.  In Fig. 5 the dimensionless pressure for the commutative and non-commutative systems are shown as a function of system size for $N=150$ particles, temperature $\tilde\beta=1$ and angular momentum $L=0$. At large system sizes the pressures agree, but they deviate strongly at smaller system size.  At system size $M=10$ the non-commutative pressure diverges, while this only happens at system size zero for the commutative case.  As we are at low temperatures here ($\tilde\beta$ is of the same order or less than the dimensionless single particle level spacing), we are in the degenerate limit.  For $L=0$ it is easy to see that the behaviour of the commutative gas should be the same as an ordinary free Fermi gas and one can quite easily estimate the behaviour of the dimensionless pressure (in two dimensions) in the degenerate limit to be $\tilde P\propto (2M+1)^{-2}$ (see \cite{pathria} p.215).  This is shown by the solid line in Fig. 5.   Naively one might expect the behaviour of the non-commutative gas to be simply that of a gas in which particles occupy a finite size that gives rise to an excluded volume, e.g., the van der Waals gas. However, it turns out not to be the case.  The behaviour of the pressure in the non-commutative case is not simply $\tilde P\propto (2M-2M_0)^{-2}$ with $M_0$ defined by the minimal area.  As is clear from Fig. 5, the pressure still behaves as $\tilde P\propto (2M+1)^{-2}$ at large system sizes, but it deviates from this behaviour at system sizes close to the minimal system size.  One can, of course, fit curves through the points in Fig. 5, but this does not really lead to new insight and we refrain from doing this.  It would be much more helpful if a simple analytic understanding, as in the commutative case, of this behaviour can be developed, but we have not been able to do it thus far. 
    
\setlength{\unitlength}{1mm}
\begin{figure}
\begin{picture}(53,53)
\put(-20, 0){\epsfig{file=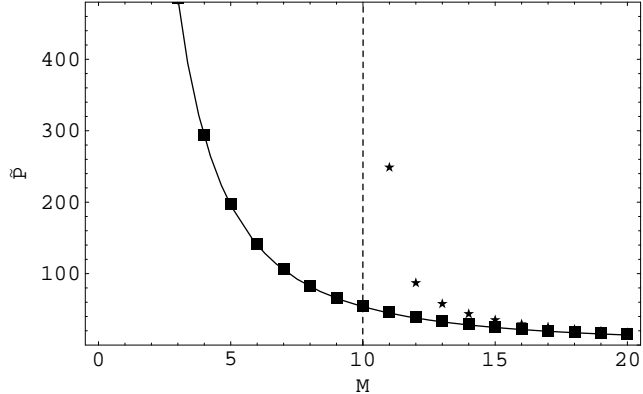, height=53mm}}
\end{picture}
\caption {Dimensionless pressure as a function of system size $M$ for the non-commutative (stars) and commutative (box) systems at temperature $\tilde\beta=1$, particle number $N=150$ and angular momentum $L=0$.  The minimal system size at which the pressure diverges is indicated by the vertical dashed line. The solid line shows the degenerate pressure for the commutative system given by $\tilde P=23\,685/(2M+1)^2$.  The proportionality constant was determined by fitting the data point at $M=0$.}
\end{figure}

The entropy of the commutative and non-commutative systems also reflects the fundamental differences at high densities. Fig. 6 shows the entropy as a function of density for a system size $M=10$ and angular momentum $L=0$.  One observes that the entropies of the two systems coincide at low densities, but that they diverge strongly at high densities and in particular that the entropy of the non-commutative system decreases.  This behaviour can easily be understood by looking at the non-commutative spectrum from which it is clear that the density of states available to the system at high densities and fixed angular momentum must decrease. Indeed, for the extremal cases discussed earlier the entropy vanishes at the critical density. 
\setlength{\unitlength}{1mm}
\begin{figure}
\begin{picture}(53,53)
\put(-20, 0){\epsfig{file=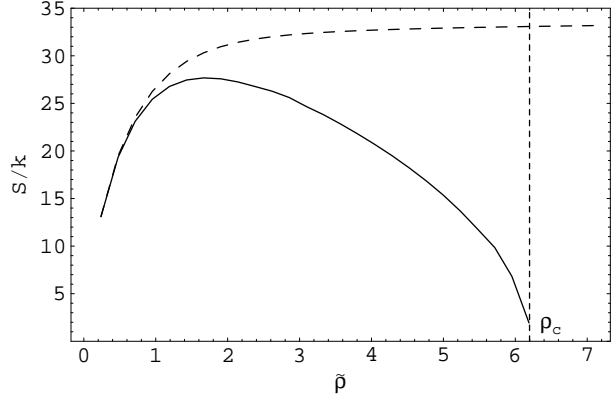, height=53mm}}
\end{picture}
\caption {Entropy in units of $k$ as a function of density for the non-commutative (solid line) and commutative (dashed line) systems at temperature $\tilde\beta=1$, system size $M=10$ and angular momentum $L=0$.  The critical density at $\tilde\beta=1$ is indicated by the vertical dashed line.}
\end{figure}

Fig. 7 shows the entropy as a function of system size for particle number $N=150$, temperature $\tilde\beta=1$ and $L=0$.  As in the case of the pressure, the behaviour of the entropy of the commutative system can again be understood easily as the degenerate limit of a free two dimensional Fermi gas.  One can then easily estimate, for a fixed number of particles, (see \cite{pathria} p. 215) that $\frac{S}{k}\propto (2M+1)$, which is clearly the behaviour exhibited in Fig. 7 as indicated by the dashed line.  The deviations at very small system size are due to finite size corrections. Note that the linear dependency of the entropy on system size is a particular feature of two dimensions and not generic.  In contrast, the behaviour of the entropy of the non-commutative system is vastly different and tends to zero at the minimal system size, indicated by the vertical dashed line.  Note that the minimal system size can be of macroscopic scale if the density is high. At large system sizes (low density as the particle number is fixed) the dependency on system size tends to that of the commutative system.  Again we lack a simple analytic understanding of this behaviour. 
\setlength{\unitlength}{1mm}
\begin{figure}
\begin{picture}(53,53)
\put(-20, 0){\epsfig{file=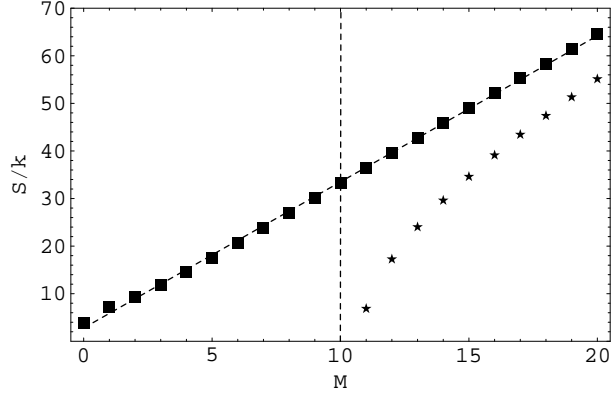, height=53mm}}
\end{picture}
\caption {Entropy in units of $k$ as a function of system size $M$ for the non-commutative (stars) and commutative (box) systems at temperature $\tilde\beta=1$, particle number $N=150$ and angular momentum $L=0$.  The dashed line depicts the behaviour of the entropy for the commutative system as derived from the degenerate limit of the free Fermi gas.  The vertical dashed line shows the minimal system size.}
\end{figure}
     
At fixed density and large system sizes, one expects intensive quantities, such as pressure and chemical potential, to be independent of system size, while extensive quantities such as entropy and energy should scale linearly with system size.  We now proceed to show that this expectation fails in the non-commutative system, even at large system sizes if the density is high enough.  The deviation from extensive behaviour is again due to the presence of a minimal system size, arising from the excluded area implied by non-commutativity. Fig. 8a shows the pressure for the non-commutative system as a function of system size $M$ at two densities $\tilde\rho=5.1$ and $\tilde\rho=4.037$, temperature $\tilde\beta=1$ and angular momentum $L=0$.  As one would expect the pressure increases sharply at the minimal system size, indicated by the dashed vertical lines, for these densities.  Here we do not compare with the commutative system as we are at fairly large densities where the commutative and non-commutative results may differ significantly and nothing can really be learned by such a comparison.  However, we do want to point out that the pressure in the commutative system only exhibits a weak dependency on the size of the system as shown in Fig.~8b for the same parameter values as in Fig. 8a (note the different scales on the vertical axis in Figs.~8a and 8b).  As mentioned before, this can be expected as pressure is an intensive quantity and should therefore be independent of system size when temperature and density are held fixed.  Indeed, in the degenerate limit the behaviour of the pressure as a function of density can easily be established to be $\tilde P\propto \tilde\rho^2$.  This is of course only true at large system size when finite size corrections are unimportant and there may be corrections at small system size as one indeed observes (see Fig.~8b).  In contrast, in the non-commutative case the sharp increase in pressure always occurs at the minimal system size, which may be large (see Fig.~2), if the density is high.  Thus this deviation is not a finite size correction, but due to the excluded area resulting from the non-commutative nature of the system.  Far enough above the minimal system size, where the excluded area is unimportant, the non-commutative system starts to behave like the commutative one and the pressure exhibits the normal intensive scaling property.
\begin{figure}[ht]
\begin{tabular}{cc}
	\epsfig{file=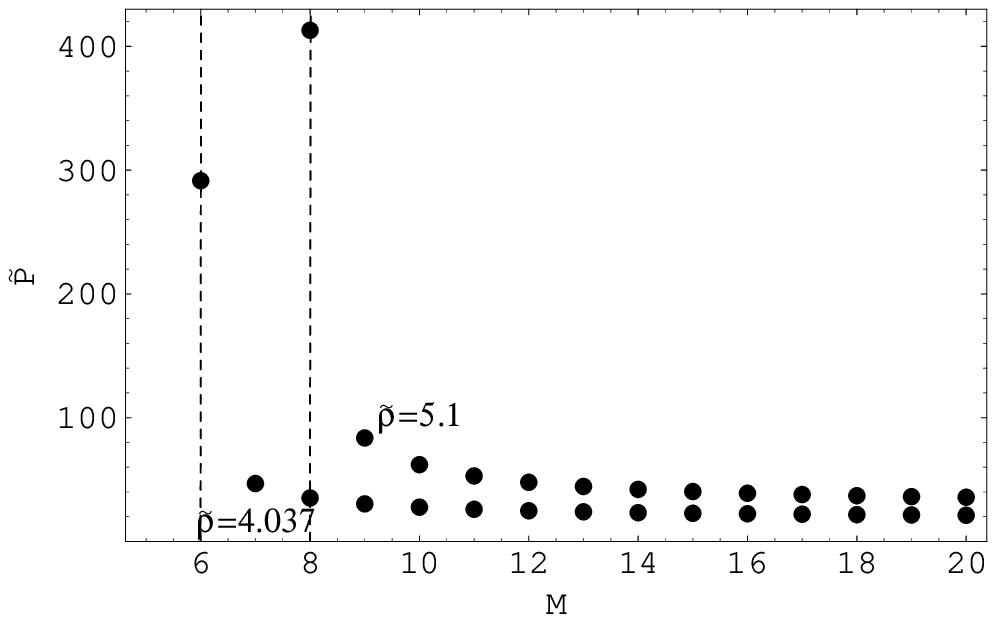,height=5cm} &
 \epsfig{file=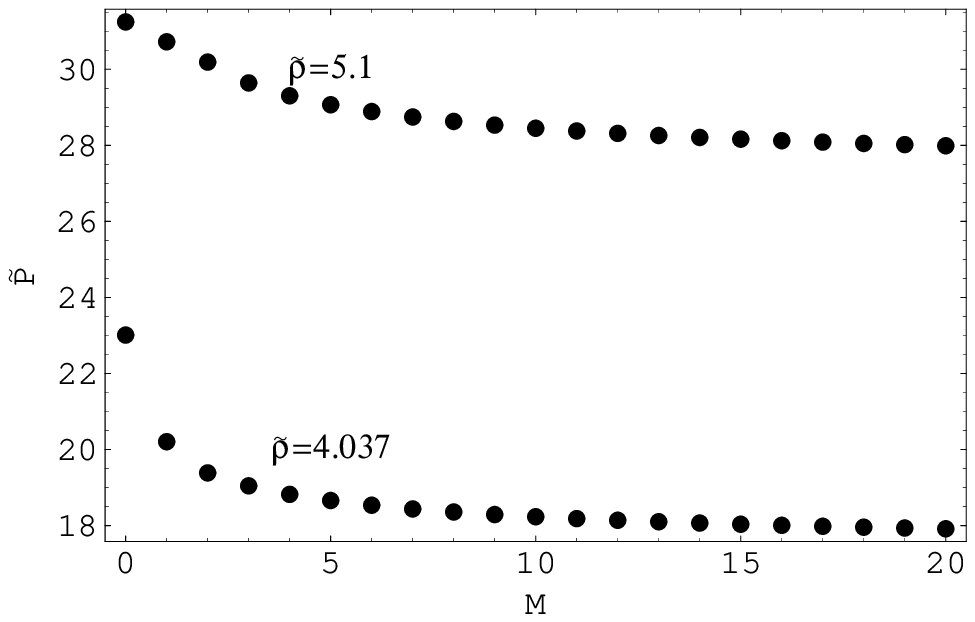,height=5cm} \\
	(a) & (b)\\
\end{tabular}
	\caption{Pressure for the non-commutative (a) and commutative (b) systems at densities $\tilde\rho=4.037$ and $\tilde\rho=5.1$, temperature $\tilde\beta=1$ and angular momentum $L=0$.  Note the difference in the scales of the vertical axis in (a) and (b).}
\end{figure}

The entropy of the non-commutative system exhibits similar scaling properties, shown in Fig. 9a, for densities $\tilde\rho=5.1$ and $\tilde\rho=4.037$, temperature $\tilde\beta=1$ and angular momentum $L=0$.  There is clearly a strong deviation from extensive scaling at the minimal system size, while extensive scaling is recovered at sizes well above the minimal system size. In contrast, the entropy of the commutative system scales extensively for virtually all system sizes as indicated in Fig. 9b for a density $\tilde\rho=5.1$ and all other parameters the same as in Fig. 9a.    \begin{figure}[ht]
\begin{tabular}{cc}
	\epsfig{file=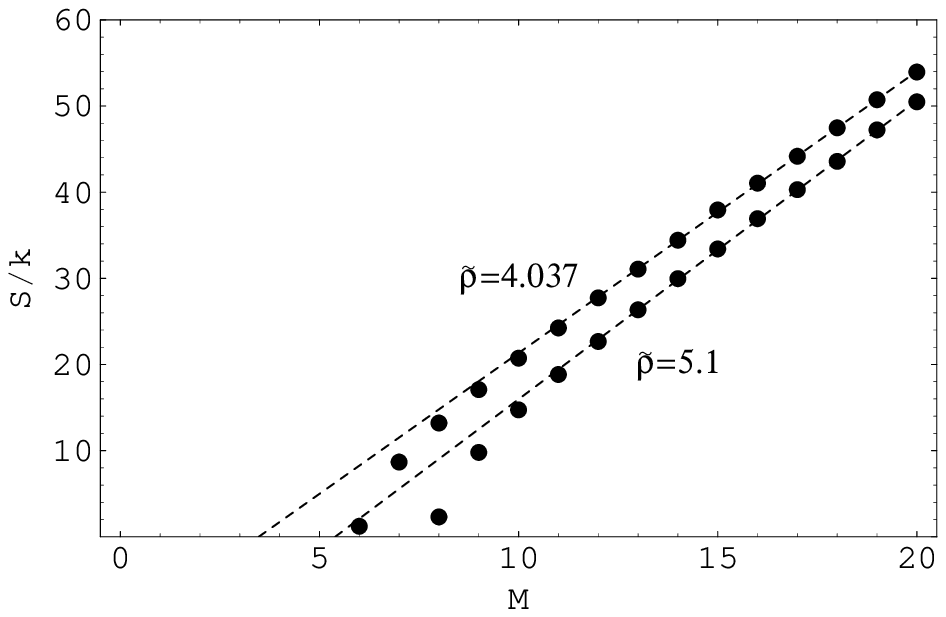,height=5cm} &
 \epsfig{file=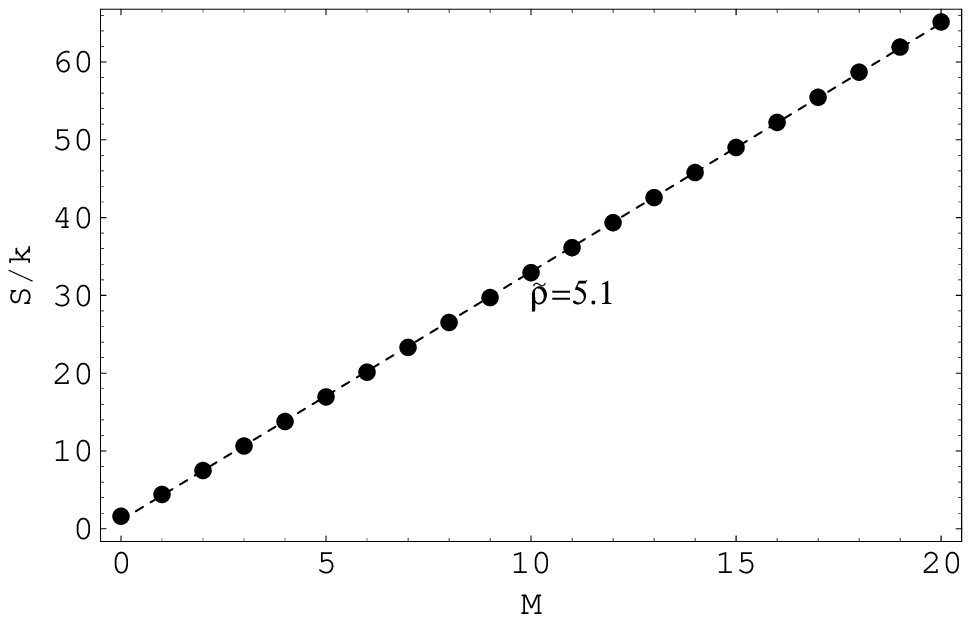,height=5cm} \\
	(a) & (b)\\
\end{tabular}
	\caption{Entropy for the non-commutative (a) and commutative (b) systems at densities $\tilde\rho=4.037$ and $\tilde\rho=5.1$, temperature $\tilde\beta=1$ and angular momentum $L=0$. The dashed lines are only intended to guide the eye and to emphasize the deviation from extensive behaviour for the non-commutative system.}
\end{figure}

\noindent{\bf Acknowledgements.}
This work was supported under a grant of the  National Research Foundation of South Africa. 
J.G. is grateful to Profs. Hendrik Geyer and Bernard Lategan for the generous support and the warm hospitality
of the Stellenbosch Institute for Advanced Study (STIAS) with the grant of a Special STIAS Fellowship
which made a recent one month stay at STIAS and NITheP possible. He acknowledges the Abdus Salam International
Centre for Theoretical Physics (ICTP, Trieste, Italy) Visiting Scholar Programme in support of
a Visiting Professorship at the UNESCO-ICMPA (Republic of Benin).
J.G.'s work is also supported by the Institut Interuniversitaire des Sciences Nucl\'eaires, and by
the Belgian Federal Office for Scientific, Technical and Cultural Affairs through
the Interuniversity Attraction Poles (IAP) P6/11.

%%%%%%%%%%%%%%%%%%%%%%%%%%%%%%%%%%%%%%%%%%%%%%%%%%%%%%%%%%%%%%%%%%%%%%%%%%%%

%%%%%%%%%%%%%%%%%%%%%%%%%%%%%%%%%%%%%%%%%%%%%%%%%%%%%%%%%%%%%%%%%%%%%%%%%%%%
%%%%%%%%%%%%%%%%%%%%%%%%%%%%%%%%%%%%%%%%%%%%%%%%%%%%%%%%%%%%%%%%%%%%%%%%%%%%
\end{document}